\documentclass[conference]{IEEEtran}

\usepackage[utf8]{inputenc}
\usepackage{graphicx}
\usepackage{amstext}
\usepackage{amsmath,amsthm}
\usepackage{amsfonts,amssymb}
\usepackage{amsmath,tabu}
\usepackage{amssymb}
\usepackage[dvips]{epsfig}
\usepackage{epsfig}
\usepackage{subfigure}
\usepackage[dvips]{graphicx}
\usepackage{multirow}
\usepackage{multirow} 
\usepackage[utf8]{inputenc}
\usepackage[english]{babel}
\usepackage{caption}
\usepackage{float}
\usepackage[nospace,noadjust]{cite}
\usepackage{algorithm,algorithmic}

\def\bPsi{{\boldsymbol{\Psi}}}

\def\b1{{\boldsymbol{1}}}
\def\c1{{\textcircled{a}}}
\def\ba{{\boldsymbol{a}}}
\def\bb{{\boldsymbol{b}}}

\def\bn{{\boldsymbol{n}}}

\def\bx{{\boldsymbol{x}}}

\def\bz{{\boldsymbol{z}}}
\def\bA{{\mathbf{A}}}

\def\bC{{\boldsymbol{C}}}

\def\bF{{\mathbf{F}}}

\def\bH{{\boldsymbol{H}}}
\def\bI{{\mathbf{I}}}

\def\bS{{\boldsymbol{S}}}

\def\bZ{{\boldsymbol{Z}}}

\newtheorem{theorem}{Theorem}[section]

\usepackage{xcolor}
\def\BibTeX{{\rm B\kern-.05em{\sc i\kern-.025em b}\kern-.08em
    T\kern-.1667em\lower.7ex\hbox{E}\kern-.125emX}}

\begin{document}

\title{Plug-and-Play Deep Energy Model for Inverse problems

\thanks{Identify applicable funding agency here. If none, delete this.}
}

\author{\IEEEauthorblockN{Jyothi Rikhab Chand}
\IEEEauthorblockA{\textit{University of Iowa} \\
Iowa city, Iowa \\
jyothi-rikhabchand@uiowa.edu}
\and
\IEEEauthorblockN{Mathews Jacob}
\IEEEauthorblockA{\textit{University of Iowa} \\
Iowa city, Iowa \\
mathews-jacob@uiowa.edu}
}

\maketitle

\begin{abstract}
We introduce a novel energy formulation for Plug- and-Play (PnP) image recovery. Traditional PnP methods that use a convolutional neural network (CNN) do not have an energy based formulation. The primary focus of this work is to introduce an energy-based PnP formulation, which relies on a CNN that learns the log of the image prior from training data. The score function is evaluated as the gradient of the energy model, which resembles a UNET with shared encoder and decoder weights. The proposed score function is thus constrained to a conservative vector field, which is the key difference with classical PnP models. The energy-based formulation offers algorithms with convergence guarantees, even when the learned score model is not a contraction. The relaxation of the contraction constraint allows the proposed model to learn more complex priors, thus offering improved performance over traditional PnP schemes. Our experiments in magnetic resonance image reconstruction demonstrates the improved performance offered by the proposed energy model over traditional PnP methods.
\end{abstract}

\section{Introduction}

The recovery of an image from a few corrupted measurements is a common problem in several applications including denoising, deblurring, and MRI image reconstruction. The maximum a posteriori (MAP) framework formulates image recovery as the minimization of a cost function that is the sum of two terms. The first term is a measure of the consistency of the image with the measurements, while the second regularization term depends on the available prior information about the image. Classical approaches rely on handcrafted priors, while recent methods use deep learned plug and play (PnP) priors that offer improved performance over traditional approaches \cite{pnpbouman,pnp_admm,ce}. 

While the empirical performance of PnP methods is superior to compressed sensing and low-rank methods, a challenge with current PnP methods is the lack of a well-defined energy formulation, typically enjoyed by classical methods. While some approaches including the RED \cite{red} introduced an energy formulation, subsequent analysis showed that many of the assumptions in RED are not satisfied by CNN based denoisers \cite{red_interp}; most PnP methods are now understood from a consensus equilibrium perspective \cite{ce}. Another limitation is the requirement that the denoiser is a contraction, which is needed for the algorithm to converge. This constraint is often enforced by spectral normalization of the individual layers \cite{sn}. In addition to restricting the possible CNN architectures, our experiments show that this constraint often restricts the CNN from learning accurate prior models, which translates to lower performance. 

The main focus of this paper is to introduce a novel plug and play energy formulation to overcome the above challenges. In particular, we use a convolution neural network (CNN) to represent the negative log density of the images. We compute the gradient of the above CNN, which models the gradient of the negative log-likelihood of the data, which is often termed as the score. The score model is pre-trained using noise score matching \cite{dsm}. The main distinction of this approach from traditional PnP models is that score function is constrained to be a conservative vector field, which satisfies the property that its line integral is independent of the specific path. The learned energy function can be used in arbitrary inverse problems by combining it with the data likelihood term to obtain the posterior distribution, which is minimized using gradient descent. 

We also introduce a novel convergence guarantee, which shows that the cost will decrease monotonically when the step-size of the algorithm is sufficiently low. The convergence guarantees are valid even when the score function is not a contraction. Traditional PnP models that learn non-conservative score functions impose an Lipschitz bound on the score function to guarantee convergence. However, this constraint may restrict the type of prior densities that can be learned. Because our network is  not required to be a contraction, it can learn more complex energy functions, which translate to improved performance. We determine the utility of the proposed scheme in the recovery of MRI data from highly undersampled measurements.


\section{Proposed Method}
\subsection{Problem Formulation}
Let us consider the problem of recovering an image $\bx \in \mathbb{R}^{m}$ from its corrupted  linear measurements $\bb \in \mathbb{R}^{n}$:
\begin{equation}\label{eq:0}
   \bA\bx+\bn = \bb
\end{equation}
where $\bA \in \mathbb{R}^{n \times m} (n \leq m) $ is a known linear transformation and $\bn \in \mathbb{N}(0,\eta^{2}\bI)$ is the additive white Gaussian noise. Then the MAP estimate of $\mathbf x$ is given by : 
\begin{equation}\label{eq:1}
	\bx^* = \arg \min_\bx \underbrace{\dfrac{1}{2\eta^{2}} \|\bA\bx-\bb\|_{2}^2}_{-\log p(\bb|\bx)} + \underbrace{\varphi(\bx)}_{-\log p_{\rm data}(\bx)}.
\end{equation}
\subsection{PnP energy model and the resulting score function}
In this paper, we use a parametric model to represent $p_{\rm data}(\bx)$:
\begin{equation}\label{prior_new}
  p_{\theta}(\bx) = \dfrac{1}{\bZ_\theta}\exp\left(-\dfrac{{\Psi}_{\theta}(\bx)}{\sigma^2}\right)
\end{equation}
where $\theta$ denote the parameters of the model and $\bZ_{\theta} $ is a normalizing constant. We model $\Psi_{\theta}(\bx): \mathbb{R}^{n} \rightarrow \mathbb{R}$ by a neural network. Here, $\sigma^{2} >0$ is the training noise variance, described later. Using the above prior in (\ref{eq:1}), the MAP objective simplifies to :
\begin{equation}\label{eq:P1}
	f(\bx)= \dfrac{1}{2\eta^{2}} \|\bA\bx-\bb\|_{2}^2 + \dfrac{1}{\sigma^{2}}\,\Psi_{\theta}(\bx)
\end{equation}

\begin{figure}[htbp]
\subfigure[Two layer network]{\includegraphics[trim={0 0cm 0 0cm},clip,width=0.48\textwidth]{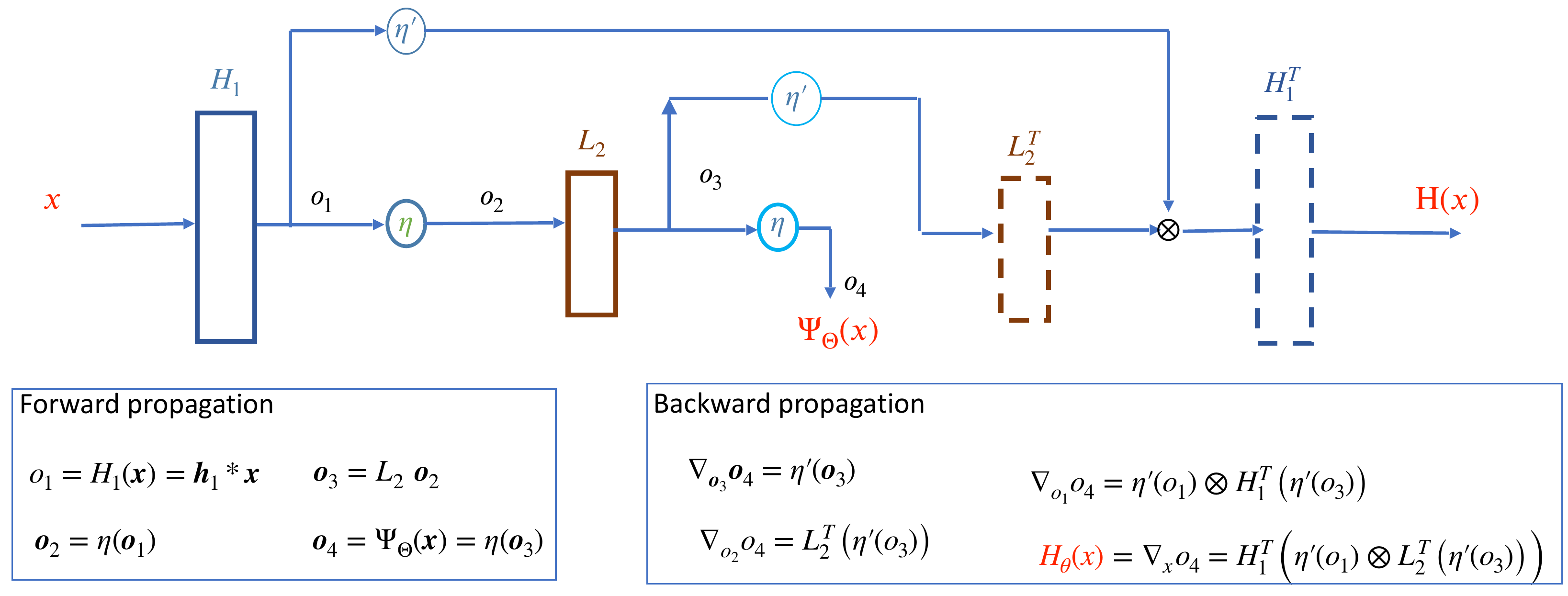}}
\subfigure[Four layer network]{\includegraphics[trim={0 9.5cm 0 3cm},clip,width=0.48\textwidth]{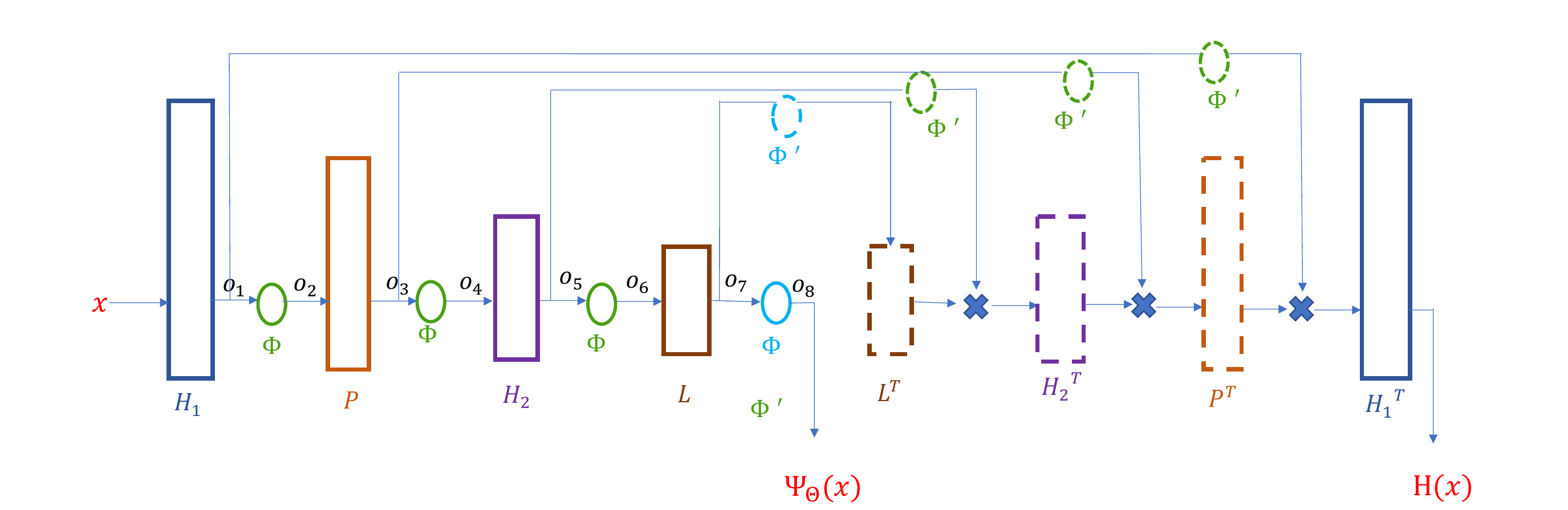}}
\caption{(a)  Illustration of a two layer energy network to realize $\Psi_{\theta}(\bx)$ and its corresponding gradient. The first and last layer are convolution and deconvolution layers with shared weights, while the second and third layers are linear and linear transpose layers with with shared weights, indicated by the same colors. The dimension of the linear layer is chosen such that $\Psi_{\Theta}(\bx)$ is a scalar.
(b) Illustration of an deeper network, where convolutional filters $H_1$ and $H_2$ denote convolution layers, while $P$ denotes a pooling layer, and $L$ denotes a linear layer. $P^T$ denotes a pooling layer. The dimensions of the layers are chosen such that $\Psi_{\Theta}(\bx)$ is a scalar. The backpropagation steps are similar to (a).}
\label{model}\vspace{-1.5em}
\end{figure}

We note that the gradient of the log prior $\nabla_{\bx}\left({-\log p_{\theta}(\bx)}\right): \mathbb R^n \rightarrow \mathbb R^n$: 
\begin{equation}\label{score}
	\nabla_{\bx}\left({-\log p_{\theta}(\bx)}\right) = \dfrac{\nabla_{\bx}\Psi_{\theta}(\bx)}{\sigma^{2}}=\dfrac{\bH(\bx)}{\sigma^{2}}
\end{equation}
is often referred to as the score function of the prior. For example, assume that $\Psi_{\theta}(\bx) = \eta(\mathbf W \bx)$, where $\mathbf W$ is a  $\mathbb{R}^{m\times m}$ vector and $\eta$ is non-linear operator, applied on each entry. Then $\nabla_{\bx} \Psi_{\theta}(\bx)$ is specified by $\mathbf W^T\eta'(\mathbf W \bx)$, which is a two layer neural network with shared weights. Fig. \ref{model} shows two different example models. From Fig.\ref{model} it is observed that when the convolutional layers use larger strides, the architecture resembles a UNET with skip connections. The output of the encoder is a scalar that yields the energy $\Psi_{\theta}(\bx)$, while the output of the decoder is the score. 
However, unlike a regular UNET, the network of the encoder and decoder are shared. In particular, the network is constrained such that $\nabla {\mathcal E}_{\theta}$ is a conservative vector field, which ensures that the resulting score network can be interpreted as the derivative of a well-defined energy function ${\mathcal E}_{\theta}(\bx)$.
\subsection{Learning the energy using denoising score matching}
Score matching \cite{dsm} aims to learn a parametric model $F_{\theta}(\bx)= \nabla_{x} \log p_{\theta}(\bx)$ from training data $\mathbf x \sim p_{\rm data}(\bx)$ such that it matches the score of the proposed distribution $\nabla_{\bx} \log p( \bx)$
\begin{equation}\label{orginal}
\theta^* = \arg \min_{\theta} \mathbb E_{\bx\sim p_{\rm data}} \| F_{\theta}( \bx) - \nabla_{ \bx} \log p_{\rm data}(\bx)\|^2.
\end{equation}
The above cost function is essentially the Fisher divergence between $p_{\rm data}(\bx)$ and $p_{\theta}(x)$. Because this cost is challenging to evaluate without the explicit knowledge of $p_{\rm data}(\bx)$, the DSM approach instead learns the score of $\tilde{\bx} = \bx + \mathbf n$, where $\mathbf n \sim \mathcal{N}(0,\sigma^{2}\bI)$. Note that $\tilde{\bx}\sim p_{\rm data}(\bx) \ast\mathcal{N}(0,\sigma^{2}\bI)$, which is a smoothed version of $p_{\rm data}(\bx)$ \cite{dsm}:
\begin{eqnarray}\nonumber
\theta^* 
 &=& \arg \min_{\theta}  \mathbb E_{\bx \sim p_{\rm data}}\mathbb E_{\bz \sim \mathcal{N}(0,\sigma^{2}\bI)}  \left\|F_{\theta} (\bx+\sigma \bz) + \frac{\bz}{\sigma}  \right\|^2\label{dsm}
\end{eqnarray}
In this approach, zero-mean Gaussian white noise of variance $\sigma$ is added to each training data point $\bx$ to yield its corresponding perturbed sample $\tilde{\bx} = \bx + \sigma\bz$ where $\bz \in \mathcal{N}(0, \bI)$. As $\sigma \rightarrow 0$, this approach minimizes the problem in (\ref{orginal}).
\\
We set $F_{\theta} = -\frac{\mathbf H_{\theta}(\bx)}{\sigma^2}$ to obtain
\begin{eqnarray}\label{dsmsimpler}
 \theta^* 
&=& \arg \min_{\theta}  \mathbb E_{p_\sigma(\bz)}  \|-\mathbf H(\bx+\sigma \bz) + \underbrace{\sigma \bz}_{\bn}  \|^2
 \end{eqnarray}

 While the approach in \eqref{dsmsimpler} is very similar to classical DSM \cite{dsm}, the main distinction is the representation of $F_{\theta}$ as the gradient vector field of $\frac{{\Psi}_{\theta}(\bx)}{\sigma^2}$. Constraining the score function as a conservative vector field allows us to use the Fundamental theorem of line integrals \cite{line} to represent the energy at any given point $x$ as the line integral:
\begin{equation}\label{LineInt}
    \log p_{\theta}(\bx) = \int_{C} F_{\theta}(\mathbf z)d\mathbf z +\log p_{\theta}(\ba) 
\end{equation}
where $\mathbf a$ is an arbitrary point and $C$ is any curve between $\ba$ and $\bx$. Because the score function is conservative, the line integral in (\ref{LineInt}) is independent of the path taken. We note that this property is not valid for arbitrary residual denoisers used in classical PnP models.
\subsection{Optimization Algorithm}
In this work, we assume the Lipschitz constant of the score network $\Psi_{\theta}$ is upper-bounded by $L$, which may be greater than one. We use the CLIP approach \cite{CLIP} to estimate the approximate constant. One may also use the product of the spectral norms to obtain an upperbound for $L$. We propose to minimize the cost function in \eqref{eq:P1} using the steepest descent algorithm: 
\begin{equation}\label{graddesc}
    \begin{array}{ll}
         \bx_{n+1}= \bx_{n}- \gamma\underbrace{\left(\dfrac{\bA^{T}(\bA\bx_{n}-\bb)}{\eta^{2}} +  \dfrac{H(\bx_{n})}{\sigma^{2}}\right)}_{\nabla f(\mathbf x_n)}.
    \end{array}
\end{equation}
We note that the fixed point of the algorithm satisfies $\nabla f(x)=0$, which correspond to the minimum of \eqref{eq:P1}.

\begin{theorem}
Let the maximum eigen-value of $\mathbf A^T\mathbf A$ is one. Then thegradient descent algorithm specified by \eqref{graddesc} will monotonically converge to a minimum of \eqref{eq:P1}, when the step-size $\gamma \leq 1/L_{\rm eq}$, where: 
\begin{equation}\label{stepsize}
    L_{\rm eq} = \dfrac{1}{\eta^{2}}+\dfrac{L}{\sigma^2}.
\end{equation}
\end{theorem}
The proof is omitted due to space constraints. Note that the Lipschitz constant of $\nabla f(x)$ is bounded by \eqref{stepsize}. The result is based on the well-known result that a Lipschitz constrained function can be upper-bounded by a quadratic function; the quadratic term depends on the Lipschitz constant. The pseudo code for the algorithm is shown below. We compute $L$ after training, which allows us to choose a step-size that can guarantee convergence. 

We note that constraining the network to learn a conservative score function ensures that the gradient magnitude decreases as one approaches the minimum, thus guaranteeing convergence. We note that the above convergence result is only applicable to energy based models; the gradient vector field learned by classical PnP methods is not constrained to be conservative. Hence, it may learn high magnitude gradients, even close to the minima. Classical methods constrain the Lipschitz constant of the score function to guarantee convergence. However, this approach often translates to networks with lower performance. 

\begin{center}
\begin{tabular}{@{}p{9cm}}
\hline
\hline
\bf{Algorithm 1: Pseudocode of EPnP GD} \\
\hline
\hline
{\bf{Input}}: Forward operator $\bA$, pre-trained denoiser $H_{\theta}(\bx_{n})$, noise variances - $\{\eta^{2}, \sigma^{2}\}$, step size $\gamma = \dfrac{1}{\dfrac{1}{\eta^{2}}+\dfrac{L}{\sigma^2}}$ \\
{\bf{Initialize}}: Set $n=0$.  Initialize $\bx_{0}$. \\ 
{\bf{Repeat}}: Given ${{\bx}_{n}}$ perform the $n+1$-th step.\\
\hspace{5mm} Compute $\bx_{n+1} =  \bx_{n}- \gamma\left(\dfrac{\bA^{T}(\bA\bx_{n}-\bb)}{\eta^{2}} +  \dfrac{H(\bx_{n})}{\sigma^{2}}\right)$\\
\hspace{10mm}$n\leftarrow n+1$\\
{\bf{until convergence}}\\
{\bf{Output}}: $\bx^{*} = \bx_{n+1}$
\\
\hline
\hline
\end{tabular}
\end{center}

\section{Experiments and Results}
\subsection{Experimental setup}
 In this paper, we evaluate the performance of the proposed method in the context of recovering parallel MRI data from highly undersampled measurements. Here, the matrix $\bA$ in (1) will be equal to $\bA=\bS\bF\bC$, where $\bS$ is the sampling matrix, $\bF$ is Fourier Transform, and $\bC$ is the coil sensitivity map. We use the publicly available dataset \cite{data}, which consists of images with a matrix size of $256 \times 232$. The training and test dataset consists of $360$ and $160$ slices, respectively. The dataset was obtained using a 12-channel head coil and CSM was pre-computed using ESPIRIT algorithm. We evaluated the proposed method using variable-density Cartesian random sampling mask with two different undersampling factors and in the presence of Gaussian noise with standard deviation $\eta=0.01$.

 \begin{figure*}[h]
\subfigure[2x acceleration]{\includegraphics[trim={7cm 6cm 5cm 5cm},clip,width=0.5\textwidth ]{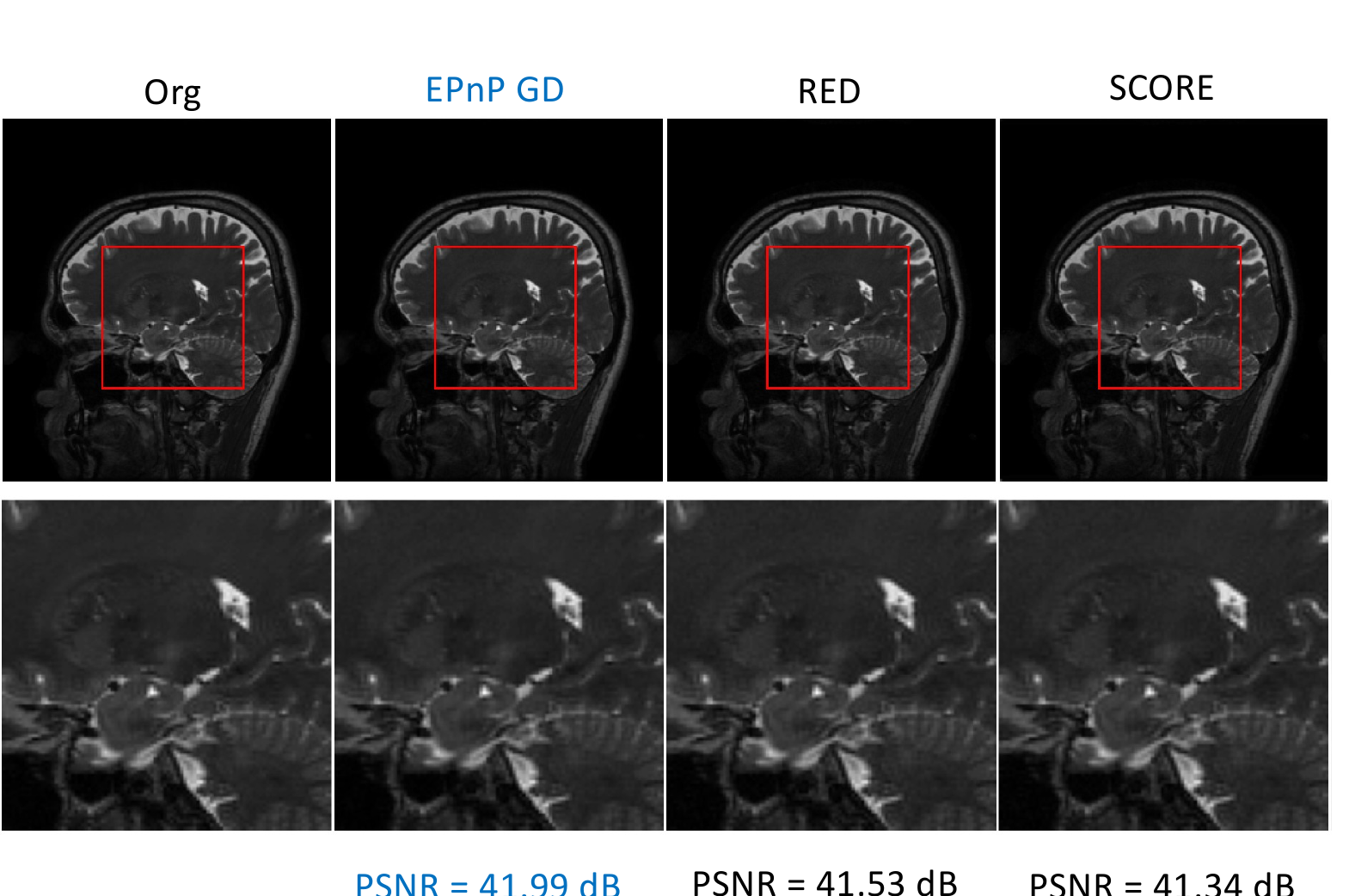}}
\subfigure[6x acceleration]{\includegraphics[trim={7cm 6cm 5cm 5cm},clip,width=0.5\textwidth ]{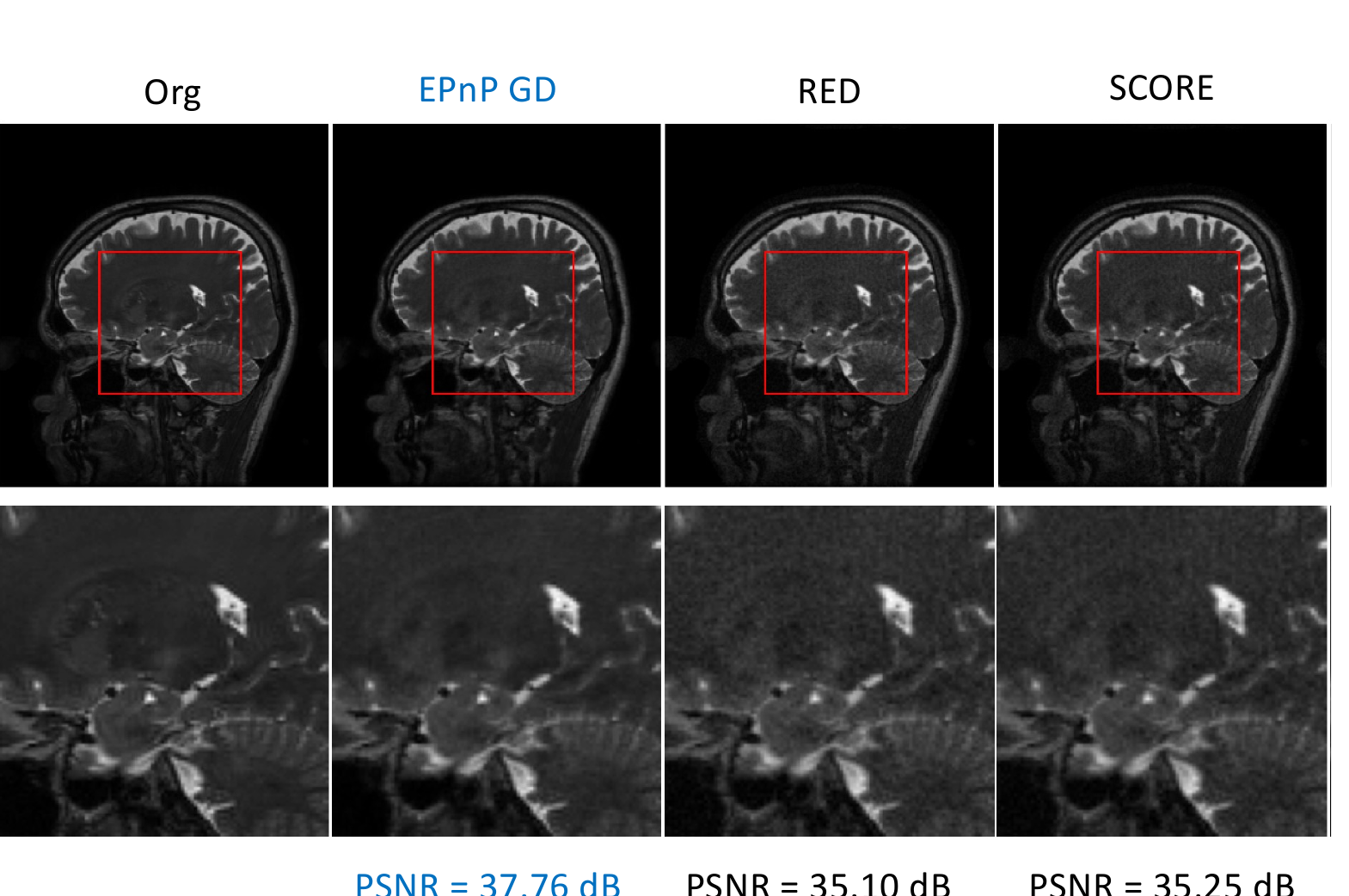}}
\vspace{-1em}\caption{Example reconstructions using the proposed EPnP algorithm, compared with RED and SCORE models, at two diffferent acceleration factors. The top row compares the reconstructed images and bottom row shows their enlarged regions for (a) two-fold (b) six-fold accelerations, respectively.}
\label{images}
\end{figure*} 

\subsection{Architecture of the networks}
We implemented the function $\bPsi_{\theta}(\bx)$ in \eqref{prior_new} using a five layer convolutional neural network that consists of $64$ channels per layer and $3 \times 3$  filters, followed by a linear layer. After each convolutional layer, ReLU activations were used. As discussed in Section II.A., the function $\bH(\bx)$ was realized using the chain rule, which guarantees the learned score to be a conservative vector field. The architecture of an example network is shown in Fig. \ref{model}. We note that the weights of the encoder and the decoder are shared. We train $\bH(\bx)$ as a noise estimator as in \eqref{dsmsimpler} with $\sigma =0.01$. The Lipschitz constant of the denoiser was estimated using the technique described in \cite{CLIP} and was used to determine the step size $\gamma$ of the steepest-descent algorithm in \eqref{graddesc}. The algorithm was run until $\dfrac{|f_{\rm{MAP}}(\bx_{n+1})-f_{\rm{MAP}}(\bx_n)|}{|f_{\rm{MAP}}(\bx_n)|}\leq 10^{-8}$ was satisfied. The proposed algorithm is refered to as EPnP-GD.

The proposed algorithm is compared with the steepest descent based RED algorithm (SD-RED) algorithm \cite{red} and a score-based model. We use a ten layer CNN with 64 channels per layers with 3x3 filters. The number of network parameters is approximately the same as the energy network used above. We use spectral normalization \cite{sn} to ensure that the networks are contractions. All the networks were trained on the training datasets described above.
\subsection{Results}
Following training, the performance of the algorithms was evaluated on the test dataset. Table. \ref{comp} shows the results for two different acceleration rates in the presence of Gaussian noise with standard deviation of $0.01$. For fair comparisons, all algorithms where initialized with $\bA^{T}\bb$. From Table \ref{comp}, we observe that EPnP GD offers improved performance compared to RED and SCORE at both accelerations. We note that RED and traditional SCORE models require the Lipschitz constant of the networks to be bounded by one, while the EPnP-GD approach does not require this constraint. We attribute the improved performance to the relaxation of the Lipschitz constraint. The improved results can also be appreciated from the reconstructions shown in Fig. \ref{images}. We note that EPnP offers reconstructions with reduced noise artifacts compared to RED and SCORE models.

\begin{table}[H]
\centering
\caption{Comparison of the proposed Energy based PnP method for two different acceleration in the presence of Gaussian noise of std $=0.01$.}
\label{comp}
\begin{tabular}{|p{2cm}|p{2cm}|p{2cm}|}
\hline
\multirow{2}{*}{Algorithm} & \multicolumn{2}{p{2cm}|}{Avg. PSNR $\pm$ std}  \\
                       & Acceleration $6$x         & Acceleration $2$x                      \\ \hline
EPnP GD                   & 37.67 $\pm$1.19              & 41.47 $\pm$ 1.18             \\ \hline

RED                     &  34.93 $\pm$ 1.71            & 40.95 $\pm$ 1.41           \\ \hline
SCORE              &34.95  $\pm$ 1.61             &40.43 $\pm$ 1.58
\\\hline 
\end{tabular}
\vspace{-1em}
\end{table}

\begin{figure}[H]
\subfigure[a]{\includegraphics[width=0.24\textwidth]{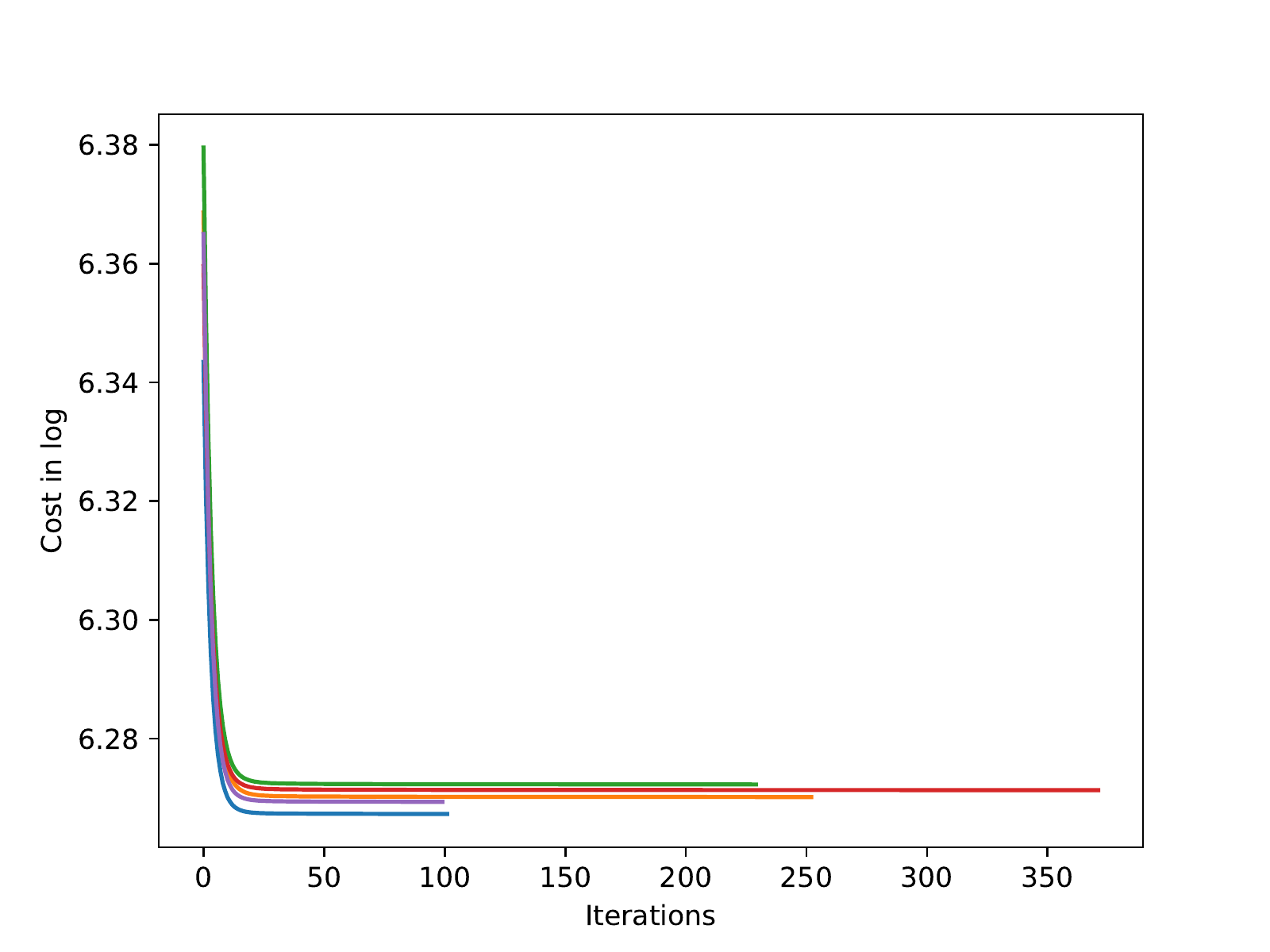}}
\subfigure[b]{\includegraphics[width=0.24\textwidth]{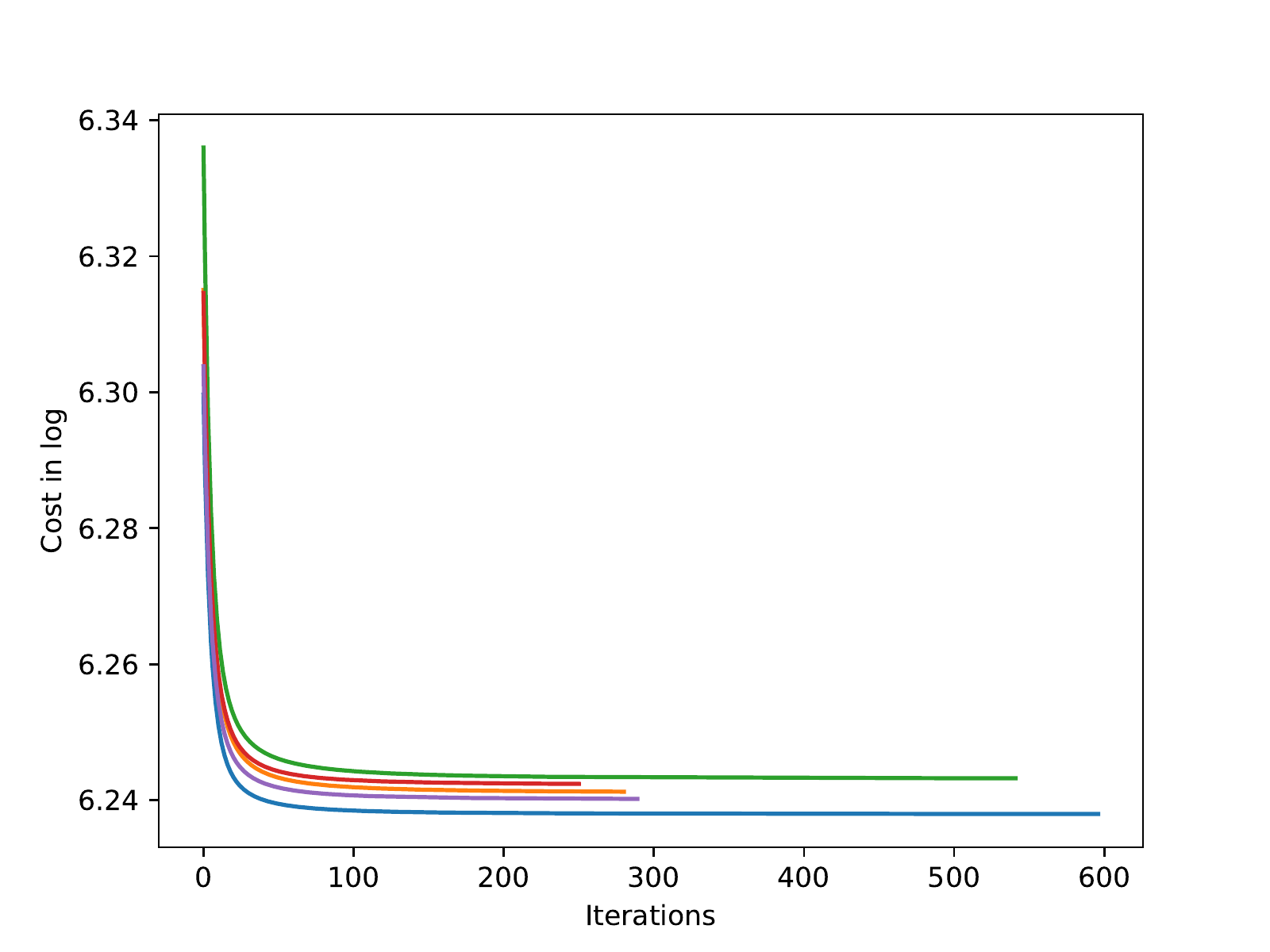}}
\caption{Convergence plot of EPnP-GD for (a) two-fold and (b) six-fold acceleration. The plots show that the algorithm converges monotonically to a minimum for different slices and acceleration factors.}
\label{cost}
\end{figure}

We note that the proposed model is associated with a well-defined cost function unlike score models. We show the convergence of the cost function in \eqref{eq:P1} for different acceleration and slices in Fig. \ref{cost}. We show the convergence plot of the EPnP-GD algorithm in Fig. \ref{cost} for ten different test slices wherein From the figure it can be observed that the EPnP-GD decreases the cost function monotonically. 

\section{Conclusion}
In this paper, we introduced a novel energy formulation for PnP framework. The proposed framework represented the log prior of the data using a deep CNN, where the gradient of the prior modeled the score function. The resulting score network resembles a U-NET whose encoder and decoder weights are shared. The parameters of the network are learned using denoising score matching. A steepest descent algorithm was used to apply the learned model to inverse problems. The proposed algorithm is guaranteed to converge monotonically to a minimum of the MAP objective, even when when the score function is not a contraction. The preliminary results in the context of MRI shows that the relaxation of the contraction constraint translates to improved performance.

\bibliographystyle{IEEEtran} 
\bibliography{ref}

\begin{thebibliography}{10}
\providecommand{\url}[1]{#1}
\csname url@samestyle\endcsname
\providecommand{\newblock}{\relax}
\providecommand{\bibinfo}[2]{#2}
\providecommand{\BIBentrySTDinterwordspacing}{\spaceskip=0pt\relax}
\providecommand{\BIBentryALTinterwordstretchfactor}{4}
\providecommand{\BIBentryALTinterwordspacing}{\spaceskip=\fontdimen2\font plus
\BIBentryALTinterwordstretchfactor\fontdimen3\font minus
  \fontdimen4\font\relax}
\providecommand{\BIBforeignlanguage}[2]{{%
\expandafter\ifx\csname l@#1\endcsname\relax
\typeout{** WARNING: IEEEtran.bst: No hyphenation pattern has been}%
\typeout{** loaded for the language `#1'. Using the pattern for}%
\typeout{** the default language instead.}%
\else
\language=\csname l@#1\endcsname
\fi
#2}}
\providecommand{\BIBdecl}{\relax}
\BIBdecl

\bibitem{pnpbouman}
S.~V. Venkatakrishnan, C.~A. Bouman, and B.~Wohlberg, ``Plug-and-play priors
  for model based reconstruction,'' in \emph{2013 IEEE Global Conference on
  Signal and Information Processing}, 2013, pp. 945--948.

\bibitem{pnp_admm}
------, ``Plug-and-play priors for model based reconstruction,'' in \emph{2013
  IEEE Global Conference on Signal and Information Processing}.\hskip 1em plus
  0.5em minus 0.4em\relax IEEE, 2013, pp. 945--948.

\bibitem{ce}
G.~T. Buzzard, S.~H. Chan, S.~Sreehari, and C.~A. Bouman, ``Plug-and-play
  unplugged: Optimization-free reconstruction using consensus equilibrium,''
  \emph{SIAM Journal on Imaging Sciences}, vol.~11, no.~3, pp. 2001--2020,
  2018.

\bibitem{red}
Y.~Romano, M.~Elad, and P.~Milanfar, ``The little engine that could:
  Regularization by denoising (red),'' \emph{SIAM Journal on Imaging Sciences},
  vol.~10, no.~4, pp. 1804--1844, 2017.

\bibitem{red_interp}
E.~T. Reehorst and P.~Schniter, ``Regularization by denoising: Clarifications
  and new interpretations,'' \emph{IEEE transactions on computational imaging},
  vol.~5, no.~1, pp. 52--67, 2018.

\bibitem{sn}
T.~Miyato, T.~Kataoka, M.~Koyama, and Y.~Yoshida, ``Spectral normalization for
  generative adversarial networks,'' \emph{arXiv preprint arXiv:1802.05957},
  2018.

\bibitem{dsm}
P.~Vincent, ``A connection between score matching and denoising autoencoders,''
  \emph{Neural computation}, vol.~23, no.~7, pp. 1661--1674, 2011.

\bibitem{line}
L.~Brugnano and F.~Iavernaro, \emph{Line integral methods for conservative
  problems}.\hskip 1em plus 0.5em minus 0.4em\relax CRC Press, 2016, vol.~13.

\bibitem{CLIP}
L.~Bungert, R.~Raab, T.~Roith, L.~Schwinn, and D.~Tenbrinck, ``Clip: Cheap
  lipschitz training of neural networks,'' in \emph{International Conference on
  Scale Space and Variational Methods in Computer Vision}.\hskip 1em plus 0.5em
  minus 0.4em\relax Springer, 2021, pp. 307--319.

\bibitem{data}
H.~K. Aggarwal, M.~P. Mani, and M.~Jacob, ``Modl: Model-based deep learning
  architecture for inverse problems,'' \emph{IEEE transactions on medical
  imaging}, vol.~38, no.~2, pp. 394--405, 2018.

\end{thebibliography}

\end{document}